# Blind Source Separation for NMR Spectra with Negative Intensity

Ryan J. McCarty*,

Department of Chemistry, University of California, Irvine,
rmccarty@uci.edu, ORCID: 0000-0002-2417-8917

Nimish Ronghe‡,

Department of Computer Science, University of California, Irvine;
ORCID: 0000-0003-1117-8165

Mandy Woo‡,

Department of Statistics, University of California, Irvine;
ORCID: 0000-0001-8866-962X

Todd M. Alam

Department of Organic Material Sciences, Sandia National Laboratories, Albuquerque, NM 87185;
ORCID: 0000-0002-1047-1231

* Correspondence: rmccarty@uci.edu; Tel.: +1-805-791-8587

‡ Nimish Ronghe and Mandy Woo contributed equally to this publication.

(Dated: 2020-02-07)

## Abstract

NMR spectral datasets, especially in systems with limited samples, can be difficult to interpret if they contain multiple chemical components (phases, polymorphs, molecules, crystals, glasses, etc...) and the possibility of overlapping resonances. In this paper, we benchmark several blind source separation techniques for analysis of NMR spectral datasets containing negative intensity. For benchmarking purposes, we generated a large synthetic datasbase of quadrupolar solid-state NMR-like spectra that model spin-lattice $T_1$ relaxation or nutation tip/flip angle experiments. Our benchmarking approach focused exclusively on the ability of blind source separation techniques to reproduce the spectra of the underlying pure components. In general, we find that FastICA (Fast Independent Component Analysis), SIMPLISMA (SIMPLe-to-use-Interactive Self-modeling Mixture Analysis), and NNMF (Non-Negative Matrix Factorization) are top-performing techniques. We demonstrate that dataset normalization approaches prior to blind source separation do not considerably improve outcomes. Within the range of noise levels studied, we did not find drastic changes to the ranking of techniques. The accuracy of FastICA and SIMPLISMA degrades quickly if excess (unreal) pure components are predicted. Our results indicate poor performance of SVD (Singular Value Decomposition) methods, and we propose alternative techniques for matrix initialization. The benchmarked techniques are also applied to real solid state NMR datasets. In general, the recommendations from the synthetic datasets agree with the recommendations and results from the real data analysis. The discussion provides some additional recommendations for spectroscopists applying blind source separation to NMR datasets, and for future benchmark studies. Applications of blind source separation to NMR datasets containing negative intensity may be especially useful for understanding complex and disordered systems with limited samples and mixtures of chemical components.

1. Introduction

Spectroscopic studies of chemical mixtures, which contain signals from multiple compounds, are challenging to study. This challenge is especially present in systems where the individual components are unknown or difficult to isolate and study individually. Examples of these systems include compounds with multiple states of order and disorder, interfaces, multiphase materials, dopants, biological systems which behave differently when isolated, and metabolomics. Nuclear Magnetic Resonance (NMR) spectroscopy is a useful tool for studying these complex mixtures, as the technique can quantitatively observe the speciation of the bulk sample, while additionally providing insight into the atomic environment, electronic structure, and ordering within the sample. A variety of techniques are available to separate and simplify NMR spectra, including physically purifying samples, selective isotopic enrichment, selective pulse sequences, multi-nuclear spectroscopic techniques, signal filters, and blind source separation techniques. Blind source separation (or blind separation) techniques [1] (such as principal component analysis or independent component analysis) are statistically based algorithms which can be used to separate components of the spectra into subsequent parts without the need to incorporate extensive information about the source signals. These techniques are useful tools but are not commonplace, possible due to barriers in implementation, such as software challenges, lack of experience, and limited literature examples.

*1.1. What is Blind Source Separation?*

Blind source separation is a broad class of approaches that separate out signals into predicted components (or parts) that can be used to recreate the input dataset. The approach is also known as component analysis, signal separation, end member separation, the cocktail party problem, unscrambling, latent variable mixture modeling, multivariate curve resolution and matrix factorization. It is a type of machine learning with both unsupervised and supervised algorithms. Blind source separation algorithms have broad applications--in social sciences, economics, music and sound, medical imaging--and are an especially popular tool for image analysis and interpreting hyperspectral images. Figure 1 visualizes the application of blind source separation to NMR spectra. "Pure components" are the perfect end members that blind source separation attempts to extract with its "predicted components". In the ideal case, predicted components are an exact match for the pure component spectra and enable chemical insight and quantification previously inaccessible. In NMR data, and often other laboratory spectroscopy techniques, spectra are scarce and relatively valuable, and the accuracy of the predicted components must be highly accurate to enable meaningful insight into the origin of the observed signal.

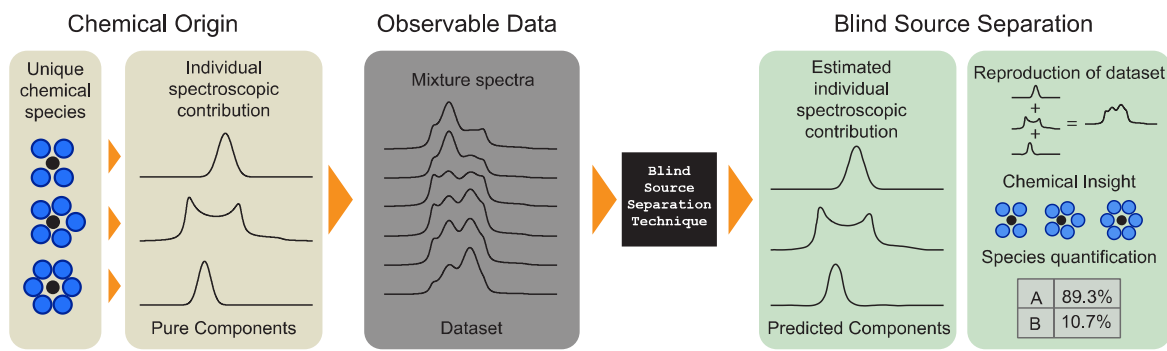

Figure 1. A flow chart indicating how unique chemical species in molecules, materials, or compounds each result in an individual spectroscopic contribution (known as pure components) which additively sum together to produce the mixture spectra. The mixture spectra (known as a dataset) are observable by NMR. The application of a blind source separation technique to the dataset results in estimated individual spectroscopic contributions (known as predicted components). Ideally, these predicted components should be able to reproduce the input dataset with minimal disagreement, enable chemical insight into the speciation, structure, and composition, and accurate quantification of the abundance of different chemical compounds.

*1.2. Previous applications of blind source separation to NMR data*

Previous applications of blind source separation to NMR data are overviewed in the review paper by Toumi et al. [2] while many of these methods for extraction of pure component spectra in a variety of applications are often captured under the Chemometrics umbrella [3,4]. The majority of applications focus on high-resolution $^1$H NMR data, typically with high signal to noise ratios and a multitude of narrow overlapping Lorentzian peaks. These studies are primarily on complex mixtures of small molecules in liquid samples. Many of the analysis methods utilize diffusion ordered spectroscopy (DOSY) to generate varying spectra which emphasize different components based on differential diffusion behavior. The DECRA (Direct Exponential Curve Resolution Algorithm) method [5–8] is most frequently used in diffusion applications, but an alternative example includes studying carbohydrate beverages using TOtal Correlation SpectroscopY (also known as TOCSY-$t_1$) encoding and diffusion encoding or $T_2$ relaxation [9]. Other examples of specific applications of blind source separation methods include analysis of $^1$H NMR data to determining the species and ratios of roasted seeds used to produce a popular drink using differences in sample composition [10] and generating metabolic profiles of saffron extracts using differences in J-coupling [11]. In less common applications, blind source separation has also been instrumental in enabling scientists to extract additional molecular level insight from NMR data. Blind source methods have also been characterizing surface structure of silica materials using variable contact time $^{29}$Si{$^1$H} cross-polarization (CP/MAS) data sets [12,13], determining the site preference of dopants in multiphase cements, ceramics, and minerals using $^{27}$Al or $^{45}$Sc NMR data sets [14], and reducing noise in $^{13}$C NMR of multiphase thermoplastics [15]. In the majority of these demonstrations use or test only a single blind source separation algorithm, raising the questions of what is the best methods for analysis of NMR spectra databases?

Benchmarking several algorithms is common in the computer science and mathematics literature, especially when presenting a new algorithm that improves accuracy or speed relative to previous work. Generalized benchmarks on all aspects of blind source separation can be challenging, as the computational time required for some matrix decomposition methods are highly sensitive to the input matrix, which depends on the type of data being studied [16].

Benchmarks of techniques relative to a specific application are less common, but in our experience, they have helped lower barriers to entry into the analysis method and enabled non-experts to more effectively and efficiently understand and apply these algorithms. Previous blind source separation benchmarks on NMR spectra focused on overlapping Gaussian peaks. Acronyms for the algorithms are detailed together in the methods and key terms and abbreviations section in this paper.

Monakhova et al. benchmarked MILCA, SNICA, JADE, RADICAL, SIMPLISMA, and MCR-ALS on $^1$H NMR datasets collected on mixtures of stock solutions and consumer products [17]. Resulting predicted components and intensities were compared using an Amari index, concentrations during synthesis, or corroborating chromatography and mass spectrometry data, to indicate that SIMPLISMA and MILCA methods demonstrated the best overall performance, SIMPLISMA and JADE produced the best quantitative analysis of concentration, and SIMPLISMA and MCR-ALS produced the best decompositions of binary mixtures [17]. Toumi et al. compared NNMF using sparse coding, and JADE, on NMR diffusion-ordered spectroscopy [18], and later benchmarked NNMF using sparse coding, JADE, and NN (the Naanaa and Nuzillard method [19]) on real $^1$H NMR spectra of sugar mixtures [2]. Using a qualitative appraisal, they determine that at low noise levels their work indicates good performance from JADE and NN, and at low and high noise, NNMF demonstrated good performance [2]. These studies are somewhat unique as they have no apparent intellectual conflicts of interest motivating results that indicate a particular technique or algorithm as being advantageous over other techniques. Cherni et al. benchmarked projected alternate least squares, soft threshold projected alternate least squares, proximal alternating linearized minimization, block-coordinate variable metric forward-backward, and wavelet-based variants on these techniques using $^1$H NMR datasets collected on mixtures of stock solutions and synthetic datasets [20]. Results were compared using an Amari index, signal to distortion ratio, signal to interference ratio, analyzed pure components, and knowledge of species quantities [20]. Cherni and co-workers drew attention to the need for reproducible and accurate peak referencing and alignment, as well as the influence of initialization of the matrix [20].

*1.3. Requirements and challenges in applying blind source separation*

Blind source separation techniques require differences in the spectra contained within the dataset. Variance can result from collecting multiple physical samples which contain different mixture compositions, by employing spectroscopic techniques which selectively emphasize or filter out different components of the spectra or both. The majority of applications of blind source separation to NMR techniques investigate exclusively positive spectra, where baseline correction and phasing produce datasets with minimal negative intensity that allows the application of blind source separation algorithms incorporating non-negative constraints. NMR techniques that produce both positive and negative spectra can present varying ratios of different components as the components change sign. Examples of these techniques include nutation experiments (which explore the sample response to changes in excitation pulse power level, also known as "tip" or "flip" angle experiments) and spin-lattice relaxation $T_1$ inversion recovery experiments. Since $T_1$ inversion and nutation are relatively straightforward experiments to preform, collecting datasets necessary for an analytical approaches that uses them is straightforward. These approaches may present attractive tools for applications limited number of samples and where users are unfamiliar with more advanced NMR pulse sequences.

*1.4. This work*

In this paper, we set out to benchmark several blind source separation techniques on NMR spectral datasets that have both positive and negative intensity. Synthetic datasets are constructed to represent the spectral response in $T_1$ inversion recovery and nutation experiments. These datasets are used for characterizing the accuracy of sixteen blind source separation techniques in estimating pure components. The insight obtained from the synthetic datasets is then applied to real experimental solid state NMR datasets. In the discussion section, we review the performance of the different blind source separation approaches and comment on their potential when applied to real datasets with components containing negative intensity. The results from the real experimental datasets motivate practical comments when applying blind source methods. We also review our approach for testing blind source separation techniques and discuss successful concepts and places for improvement.

2. Materials and Methods

Our methodological approach can be summarized into the following steps: we created mixture NMR spectral datasets, performed blind source separation using several algorithms on these datasets, and then appraised the accuracy of the resulting predicted components by comparing them to the pure components which were used to create the mixtures. We then applied these same blind source separation algorithms to experimental NMR data. All synthetic and real NMR data used in this study were in the frequency domain (i.e. the horizontal axis is the isotropic chemical shift δ_iso in ppm and the vertical axis is signal intensity in arbitrary units). Several computer languages were used during software development, but all reported work was written in python [21] using the intel distribution for python [22] version 3.6 or 3.7 on a variety of 2012 to 2018 era personal desktop or laptop computers running Windows 10, macOS, or distributions of Linux. In addition to native python libraries, we used NumPy 1.17.2 [23], Numba 0.39.0 [24], MatPlotLib 3.1.2 [25] and Pandas 0.23.0 [26] during development and production. This project also made use of Microsoft Word, Adobe Illustrator CS3, LaTeX, Google Chrome, Microsoft Edge, Mendeley, and extensive use of information and approaches presented on forums and discussion boards. Code used in this project will be publicly available at https://doi.org/10.25351/V3WC79 following the publication of a related paper, or by contacting the corresponding author (R.M.)

*2.1. Generated synthetic NMR-like datasets*

To make the synthetic mixture NMR datasets we generated "pure" components and then created additive mixtures of these pure components. Each set of mixture spectra was collected into a dataset, and blind source techniques were tested on the datasets.

We generated a database of "pure" components that represent a windowed spectrum focused on the central transition of a quadrupolar resonance. The pure components were generated using functions from the open-source NMR analysis software ssNAKE [27]. The database contains 32,000 spectra, each with 1024 points, 10,000 Hz spectra width (sweep width), a Lamour observe frequency of 100 MHz, nuclear spin of $I = 3/2$, 10 kHz spinning speed, varied quadrupolar coupling constant (cQ = 0 to 4 MHz, split into 40 equally spaced steps), varied quadrupolar asymmetry parameter ($\eta = 0$ to 1, split into 10 steps), and isotropic peak positions ($\delta_{iso}$) arranged in 10 steps across the central 7,500 Hz within the 10,000 Hz spectrum window. Gaussian smoothing was set in ssNAKE to a value of $2^n$, where n was an integer from 3 to 10, resulting in unnoticeably broadened spectra at n = 3 and broad peaks extending beyond the spectrum window for n = 10. The resulting pure components presented a broad range of peak shapes, including narrow Gaussian peaks, easily recognizable quadrupolar peaks with sharp horns, and difficult to distinguish broad features.

We created 720 mixture datasets. Each mixture dataset contains 20 spectra. Each dataset was created with 4, 6, or a randomly selected value between 2 to 10, pure components. Pure components were selected from the previously described database of pure components using reservoir sampling [28]. Mixture datasets were made with no noise, or one of 5 levels of noise (0.0001, 0.000178, 0.000316, 0.000562, and 0.001). In practice, this resulted in spectra with an approximate ratio of the total signal intensity to the absolute noise intensity of ~ 24:0, 24:1, 21:1, 15:1, 13:1, 10:1. For each set of variables tested, we created 20 datasets which vary due to the random selection of pure components, number of pure components, intensity of the pure components with $T_1$ relaxation times below as described below, and noise. The exact same datasets were used for testing of the different blind source algorithms. The pure components used in each dataset were recorded separately for later comparison. The intensities of the pure components were selected to represent an inversion recovery mixture or a partially selective nutation mixture.

In the spin-lattice $T_1$ inversion recovery datasets, the intensity values follow the ideal inversion recovery equation (equation 1), where each spectrum in the dataset represented a different recovery time $\tau$.

$$\mathbf{intensity}_{[i]} = A - 2A\, e^{\left(\frac{\tau}{T_1}\right)} \tag{1}$$

In equation 1, $i$ is the index for each pure component which is assigned a $T_1$ value, intensity is the intensity of the pure component at a given time $\tau$, A is the fully relaxed intensity of the pure component, $e$ is the exponential function to the base 2.71828, $\tau$ (tau) is the magnetization recovery time between an initial magnetization inverting 180 degree pulse, and a subsequent 90 pulse, and $T_1$ is the nuclear spin-lattice relaxation time of component $i$. 20 mixture spectra were included in each dataset at equally spaced tau values, and the final $\tau$ value was always selected such that all components were 98.5% full intensity or greater. A was randomly selected such that no component had less than 20% of the intensity of the largest component. $T_1$ was randomly selected between 0.5 and 2 for each $i$ (pure component).

In the nutation NMR datasets, the intensity values were determined using equation 2, which assumes that the first observation was at an optimal tip/flip angle and additional spectra result from further increases in the the power level.

$$\textit{intensity} = A \times \text{cosine}(2\pi \times f \times \text{pulse}) \qquad (2)$$

In equation 2, *intensity* is the intensity of a specific pure component at a given "pulse", A is the maximum possible intensity of the pure component, $\pi$ is Archimedes' Constant (3.1415...), $f$ is the ordinary frequency of a specific component, and **pulse** is an array of values with a length equal to the number of spectra in the dataset and values from 0 to 1. One of the pure components always had an ordinal frequency value of 0.50 and the rest were selected randomly between 0.50 to 0.75, an approximation for differences in peak intensity due to selective excitation. The maximum possible intensity (A) was randomly selected between 0 and 1 for each pure component. This is a simplistic model that does not accurately model nutation experiments; however, it provides differences among the components which can be analogous to nutation data. We expect it to give a good indication of the performance of blind source separation techniques on nutation datasets.

*2.2. Experimental NMR dataset*

Solid state [1]H magic angle spinning (MAS) NMR spectra were obtained using a Bruker Avance III spectrometer at a proton observation frequency of 600.1 MHz on a 2.5 mm broadband MAS probe spinning at 30 kHz. A rotor-synchronized Hahn echo pulse sequence with a 5 s recycle delay, and a $\pi/2$ pulse length of 2.5 µs was used. Data consisted of 1K complex zero-filled to 4K, Fourier transformed followed by baseline correction. An inversion recovery pulse sequence incorporating the Hahn Echo, typically used for the determination of spin-lattice relaxation times ($T_1$), was used to produce the $T_1$ relaxation-modulated spectral data set.

An existing nutation dataset of [27]Al (I =5/2) NMR at 14.1 Tesla of yttrium aluminum garnet doped with 0.9% Tm [29] was used. Spinning speed was kept at 20.0 kHz, and pulse width was varied from 2 to 6 µs, where 4.25 µs was approximately equal to the 180 ° tip angle of several paramagnetic peaks resulting from $AlO_6$ sites in the material. In the original publication, this dataset was used to distinguish the diamagnetic $AlO_6$ peak and associated paramagnetic shifts from overlapping $AlO_4$ sites which have a different response to the pulse width. A windowed section 10200 points wide was used for testing of the blind source methods.

*2.3. Statistical methods and algorithms*

We benchmarked the following blind source separation algorithms. Readers seeking additional explanations or background on the blind source separation techniques used are encouraged to search for tutorials and discussion online. Online materials related to machine learning and blind source separation are, in our experience, better reviewed, more frequently updated, less one-sided, and overall more accessible than peer-reviewed academic materials that would typically be listed here.

SVD (Singular Value Decomposition) [30,31] as implemented in NumPy [23]. Truncated SVD as implemented scikit-learn [32] using the fast randomized solver [16] (*Truncated SVD-randomized*) or the eigenvalue solver from ARPACK (ARnoldi PACKage, https://www.caam.rice.edu/software/ARPACK/) (*Truncated SVD-arpack*) used in Sci-Py [33].

PCA (Principal Component Analysis) [34,35] as implemented in scikit-learn [32]. Sparse PCA [36] as implemented in scikit-learn [32] with both the least angle regression (*Sparse PCA-lars*) and coordinate descent (*Sparse PCA-cd*) methods. Incremental PCA [37], a PCA approach incorporating the Sequential Karhunen–Loeve algorithm of Levy and Lindenbaumt [38]. TGA (Trimmed Grassmann Average) [39], a robust PCA (as implemented by Jiyuan (Glenn) Qian, a derivative of the MATLAB version by Hauberg [39]. PARAFAC [40,41] initialized using a random (*PARAFAC-random*) or SVD (*PARAFAC-svd*) starting matrices as implemented in TensorLy [42], a derivative the work by Rasmus Bro [43].

Fast ICA (Independent Component Analysis) (Hyvärinen 1999) as implemented in scikit-learn [32]. MILCA (Mutual Information Least dependent Component Analysis) [44,45] as provided by github user nordavinden (https://github.com/nordavinden/mikstur) using least angle regression (*MILCA-lars*) or coordinate descent (*MILCA-cd*). JADE (Joint Approximate Diagonalization of Eigenmatrices) [46] as implemented by Gabriel J.L. Beckers [47] a derivative of the MATLAB version available by contacting Jean François Cardoso [48].

VCA (Vertex Component Analysis) [49] as implemented by Adrien Lagrange [50] a derivative of the MATLAB version provided by Nascimento and Dias.

NNMF (Non-Negative Matrix Factorization, also frequently abbreviated as NMF) [51] as implemented in scikit-learn [32] using coordinate descent [52] and initialized using random matrices (*NNMF-random*), non-negative double singular value decomposition [53] with zero values: as zeros (*NNMF-nndsvd*), replaced with average point value of the input dataset (*NNMF-nndsvda*), or replaced with a random very small positive values (*NNMF-nndsvdar*). Since our datasets intentionally break the non-negative requirement necessary for input data using this technique, spectra with negative values were inverted if they contained more negative intensity than positive intensity, offset from the real baseline by a positive value which ensures no negative intensity, or both.

SOBI (Second Order Blind Identification) [54] as implemented by David Rigie [55].

MCR (Multivariate Curve Resolution) [56] as implemented by Charles H. Camp [57,58], using ordinary least squares [32], sparse coefficients [32] non-negative least squares (*MCR-NNLS*) using the Karush-Kuhn-Tucker conditions [33,59], ridge regression [32] (*-ridge*), and the gaussian method (*-Gauss*) found in the run/test code [58].

SIMPLISMA (SIMPLe-to-use-Interactive Self-modeling Mixture Analysis) [60–62] as implemented by Mandy Woo and Ryan McCarty, a derivative of the MATLAB code written by Willem Windig. The "offset" values were set to 0, 2, 8, 12, and 15 (resulting in the corresponding *SIMPLISMA-offset#*).

Spectra were normalized before blind source separation with three different approaches, as raw mixture spectra (as generated, equivalent to spectra as collected by an instrument), mixture spectra normalized to equal peak height, and mixture spectra normalized to have equal absolute intensity. Our implementations of SVD, JADE, incremental PCA, sparse PCA, TGA, and MILCA, were not influenced by normalization.

Most techniques require a user inputted number of components to predict (typically unknown). For these techniques, we set the predicted component number equal to the true number of components used in the synthesis of the dataset, or to 1 and 2 components below the true value, or to 1, 2, 3 and 4 components above the true number of components. Some techniques predict an internally decided number of components (typically a number equal to the number of input spectra). For these techniques, we exported the top components for the desired number of predicted components and discarded the remaining components prior to error analysis.

## 2.4. Quantifying performance

For each technique, we appraised the accuracy (goodness of fit) by comparing predicted component spectra of each dataset to the pure component spectra which were used to create the dataset. Components predicted in the different algorithms are not indexed, and therefore must be compared with each of the possible pure components to determine the best match. Furthermore, each predicted component has an algorithm determined vertical intensity scale along with a possible intensity offset. To compare predicted and pure components, the predicted component was fit to the pure component by optimizing a vertical multiplier M and an additive offset value B (see equation 3) by minimizing the resulting "lack-of-fit sum of squares error" resulting from the difference between the predicted and pure components. With our synthetic data, the statistical "pure-error sum of squares" is so small it can be neglected. The Nelder-Mead minimization approach [63] was used as implemented in SciPy [33]. The minimized total squared error was recorded for every possible match.

$$\text{lack-of-fit}_{[n]} = \sum_i \left| \text{predicted\_intensity}_{[i]} - \left( B + \text{pure\_intensity}_{[i]} \times M \right) \right|^2 \quad (3)$$

In equation 3, lack-of-fit is the sum of squares due to lack of fit, $n$ identifies the specific predicted to pure component match, $i$ is the index for each point in the spectrum, B is a signed offset, M is a signed vertical multiplier, predicted_intensity is the list of intensity values for each point of the predicted component, pure_intensity is the list of intensity values for each point of the pure components. All intensity values of the pure component are positive, and it was not uncommon for the M multiplier to be used to invert negative predicted components.

To assign a specific predicted component to a pure component, the ensemble of matches must be considered. Every possible ensemble of predicted versus pure component assignments were considered. A key constraint we imposed is that a predicted component can only be assigned to a single pure component, and each pure component can only be assigned a single predicted component. For selecting this match, the inverse total squared error (see equation 4) for the ensemble was summed for each ensemble. The ensemble with the largest inverse error was selected as the most realistic predicted component to pure components match.

$$\text{ensemble\_inverse\_error} = \sum_n \frac{1}{\text{lack-of-fit}_{[n]}} \quad (4)$$

Where $\Sigma$ (sigma) implies the sum of the inverse lack-of-fit of the ensemble, lack-of-fit is the total squared errors described in equation 3 for a specific predicted to pure component match, and $n$ denotes each match in the ensemble. Selecting ensembles using the ensemble inverse error prevents predicted components that have very large errors from stealing the match of a very close predicted component to pure component. If there were more predicted components than pure components, the best predicted to pure matches were determined, and excess predicted components were discarded, resulting in no contribution to the ensemble_inverse_error. In this manner the predicted components are indexed the specific pure components.

The best ensemble of matches was calculated for each prediction of each technique (15), with each test variable and normalization (1 to 15 depending on the technique), on each dataset (20). From this data, we determined standard measures of center, min, max, and variance for the resulting best ensemble total squared errors for each technique. If a given technique failed on a specific dataset (i.e. it could not produce any predicted components for the specific settings), its contribution to the errors was not accounted for.

3. Results

*3.1. Results from synthetic datasets*

We tested the blind source separation algorithms on the constructed synthetic datasets. Table 1 presents the accuracy of these techniques (in a *mean (minimum, maximum)* format) across our 720 test datasets and reports an approximate "runtime factor" (the most frequent magnitude of the algorithm runtime in seconds) of the implementations of the algorithms we used. The overall accuracy indicates the broad most accurate technique; however, as we discuss later, the minimum values and range of accuracy of the technique should be given consideration. Furthermore, the data reported in Table 1 assumes that the number of components in the dataset is known exactly. Figure 2 plots the nutation and inversion datasets that contributed to Table 1 separately. The larger range of the inversion dataset is visible in the figure, as well as the generally similar means of the two datasets.

Table 1. Synthetic Dataset mean squared errors: *mean (minimum, maximum)* and runtime factor for each technique.

| Technique[1] | No Normalization | Peak Normalization | Area Normalization | Runtime Factor[2] |
|---|---|---|---|---|
| FastICA | 0.24 (2.22E-04, 4.05) | 0.23 (3.40E-04, 4.05) | 0.26 (2.22E-04, 7.36) | -1 |
| SIMPLISMA-offset12 | 0.32 (2.16E-08, 6.79) | 0.43 (1.34E-03, 21.33) | 0.61 (3.81E-04, 17.2) | 0 |
| SIMPLISMA-offset8 | 0.32 (1.10E-08, 6.99) | 0.44 (1.34E-03, 21.33) | 0.56 (3.81E-04, 17.2) | 0 |
| NNMF-nndsvd | 0.32 (2.07E-05, 5.7) | 0.33 (2.07E-05, 5.67) | 0.32 (2.07E-05, 5.68) | -2 |
| NNMF-nndsvdar | 0.33 (2.07E-05, 5.62) | 0.32 (2.07E-05, 5.63) | 0.33 (2.07E-05, 5.68) | -2 |
| SIMPLISMA-offset15 | 0.32 (2.16E-08, 6.46) | 0.44 (1.34E-03, 21.33) | 0.6 (3.81E-04, 16.43) | 0 |
| SIMPLISMA-offset2 | 0.36 (8.06E-09, 13.55) | 0.48 (1.34E-03, 21.11) | 0.56 (3.81E-04, 17.2) | 0 |
| NNMF-nndsvda | 0.37 (2.07E-05, 4) | 0.39 (2.07E-05, 5.95) | 0.38 (2.07E-05, 5.61) | -2 |
| NNMF-random | 0.41 (9.92E-05, 6.43) | 0.4 (9.12E-05, 6.94) | 0.39 (9.20E-05, 7.67) | -2 |
| TGA | 0.43 (1.21E-03, 7.64) | n.a.[3] | n.a.[3] | 0 |
| SIMPLISMA-offset0 | 0.44 (3.64E-09, 10.15) | 0.49 (1.34E-03, 21.11) | 0.52 (3.81E-04, 17.2) | 0 |
| VCA | 0.44 (2.10E-03, 8.79) | 0.45 (2.10E-03, 9.43) | 0.46 (2.10E-03, 11.63) | -1 |
| JADE | 0.45 (1.27E-03, 9.52) | n.a.[3] | n.a.[3] | -1 |
| PARAFAC-random | 0.45 (8.27E-04, 10.66) | 0.46 (1.26E-03, 8.68) | 0.45 (6.99E-04, 8.6) | -2 |
| MCR-NNLS-random | 0.48 (1.34E-03, 17.51) | 0.54 (1.35E-03, 18.88) | 0.49 (1.34E-03, 21.25) | 1 |
| MCR-AR-Gauss-random | 0.59 (2.46E-03, 42.27) | 0.48 (2.46E-03, 42.27) | 0.68 (2.46E-03, 43.59) | 1 |
| MILCA-cd | 0.5 (5.56E-05, 21.58) | n.a.[3] | n.a.[3] | 1 |
| MILCA-lars | 0.5 (5.56E-05, 21.58) | n.a.[3] | n.a.[3] | 1 |
| MCR-NNLS | 0.51 (1.34E-03, 17.51) | 0.5 (1.34E-03, 17.51) | 0.51 (1.34E-03, 17.51) | 1 |
| MCR-AR-Gauss | 0.56 (2.46E-03, 42.27) | 0.76 (2.46E-03, 50) | 0.93 (2.46E-03, 43.6) | 1 |
| MCR-ALS-random | 0.7 (2.57E-03, 22.9) | 0.81 (2.08E-03, 27.62) | 0.72 (2.91E-03, 23.06) | 1 |
| MCR-ALS | 0.7 (1.34E-03, 14.43) | 0.74 (1.34E-03, 23.87) | 0.71 (1.34E-03, 17.51) | 1 |
| Truncated SVD-random | 0.77 (1.09E-03, 43.71) | 0.77 (1.09E-03, 43.71) | 0.77 (1.09E-03, 43.71) | -3 |
| PARAFAC-svd | 0.77 (1.09E-03, 43.71) | 0.77 (1.09E-03, 43.71) | 0.77 (1.09E-03, 43.71) | -2 |
| Truncated SVD-arpack | 0.78 (1.09E-03, 43.71) | 0.78 (1.09E-03, 43.71) | 0.78 (1.09E-03, 43.71) | -3 |
| PCA | 0.8 (1.10E-03, 43.12) | 0.8 (1.10E-03, 43.12) | 0.8 (1.10E-03, 43.12) | -2 |
| MCR-AR-Ridge | 0.82 (1.34E-03, 23.15) | 0.85 (1.34E-03, 23.01) | 0.8 (1.34E-03, 17.5) | 1 |
| MCR-AR-Ridge-random | 0.83 (1.45E-03, 23.64) | 0.81 (1.42E-03, 23.1) | 0.81 (1.45E-03, 17.93) | 1 |
| Incremental PCA | 0.81 (1.20E-03, 43.12) | n.a.[3] | n.a.[3] | -3 |
| SOBI | 0.89 (3.32E-04, 43) | 0.91 (3.32E-04, 42.1) | 0.9 (3.32E-04, 42.56) | 1 |
| Sparse PCA-cd | 0.9 (4.98E-03, 24.19) | n.a.[3] | n.a.[3] | -2 |
| Sparse PCA-lars | 0.9 (4.98E-03, 24.19) | n.a.[3] | n.a.[3] | -2 |
| SVD | 1.09 (2.51E-03, 43.42) | n.a.[3] | n.a.[3] | -2 |

For the three normalization types, the mean of the mean squared error value is given, and in parenthesis, the minimum error and maximum error. Reported values are for blind source separation of the exact number of components as contained in the dataset. Green highlighting is used to draw the reader's attention to which normalization approach provides the lowest mean squared error.

[1] Techniques and abbreviations are described in the methods section.

[2] The Runtime Factor is the magnitude of the most frequent runtime on our datasets with 10,000 points, 20 individual spectra and 2 or 10 components. The time factor values should be only used for context; see "Algorithm run times" in the discussion section for additional commentary on these values.

[3] The algorithm contains internal normalization resulting in identical results regardless of normalization type.

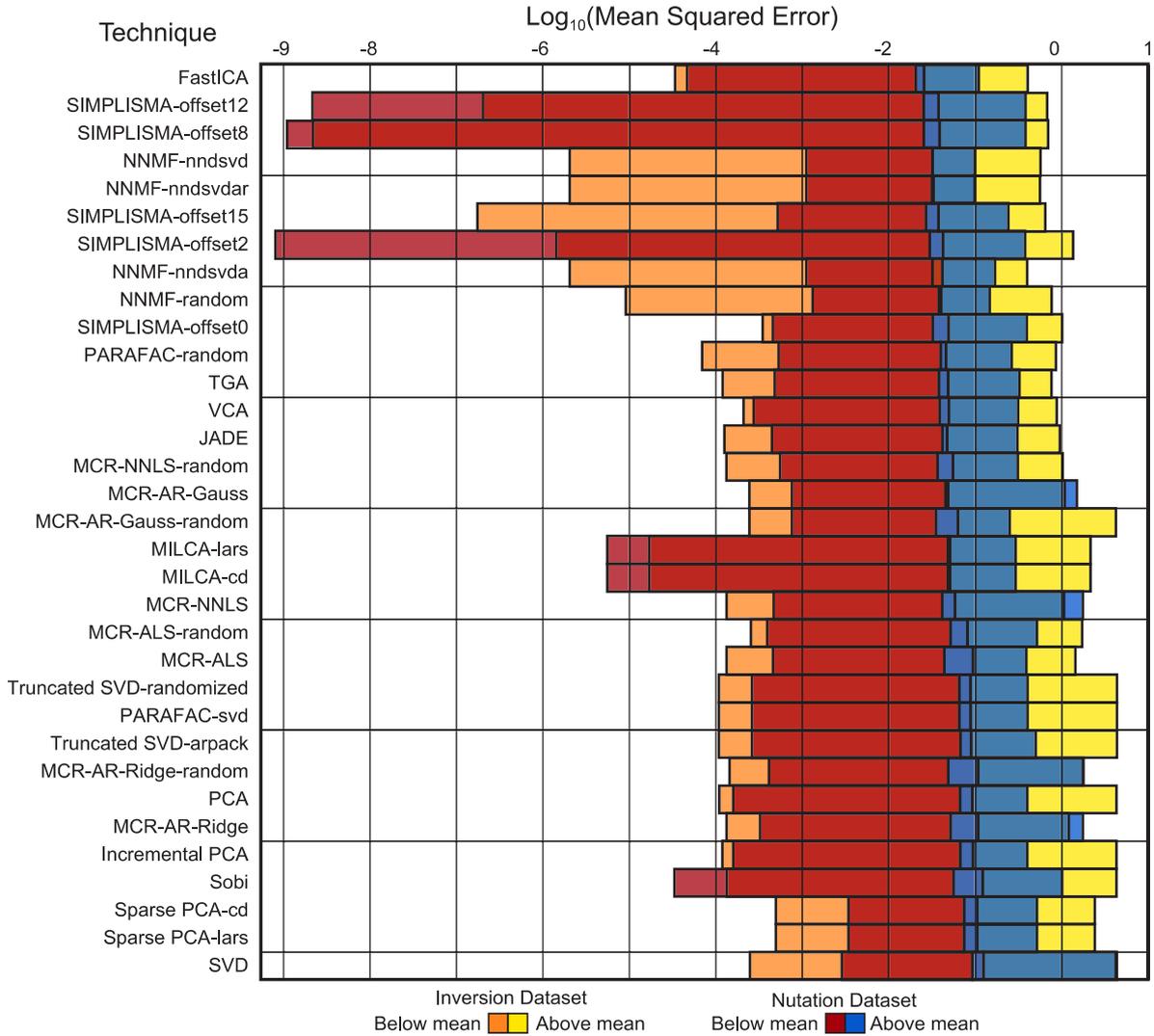

Figure 2. The minimum, maximum and mean of the mean squared errors from the inversion and nutation datasets plotted on a logarithmic scale base 10. The methods are arranged in the order from Table 1. The nutation dataset (plotted on top) is slightly transparent so that the underlying inversion dataset can be seen as well. In general, the nutation data spans a smaller range of errors and contributes a lower mean.

The number of components in a dataset can be challenging to determine. There are methods and techniques for estimating the number of components in a dataset, and although we do not use or benchmark these techniques, our study does provide insight into changes in accuracy due to predicting excess components. Table 2 summarizes the change in mean squared error relative to an exact number of components as additional components are predicted. Since our quantification of performance only determines the accuracy for the best matching components, values close to 1 indicate that predicting additional components does not degrade the accuracy of the previously predicted components. Increased values over 1 indicate that intensity relating to pure components must be mixed or separated into additional components which are discarded. Occasionally, PARAFAC and MCR with the gaussian setting were unable to converge to a solution and did not predict a component.

Table 2. Fractional increase in relative mean squared error with the prediction of additional components in the synthetic Dataset.

| Technique | Exact | + 1 | + 2 | + 3 | + 4 |
| --- | --- | --- | --- | --- | --- |
| VCA | 1 | 0.96 | 0.95 | 0.96 | 0.96 |
| MCR-NNLS | 1 | 0.96 | 0.99 | 1.00 | 0.98 |
| PCA | 1 | 1.00 | 1.00 | 1.00 | 1.00 |
| Incremental PCA | 1 | 1.00 | 1.00 | 1.00 | 1.00 |
| Sparse PCA | 1 | 1 | 1 | 1 | 1 |
| TGA | 1 | 1 | 1 | 1 | 1 |
| PARAFAC | 1 | 1.01 | 1.02 | 1.01 | 1.01 |
| JADE | 1 | 1.02 | 1.02 | 1.02 | 1.04 |
| NNMF | 1 | 1.00 | 1.01 | 1.04 | 1.04 |
| MCR-AR-Ridge | 1 | 1.04 | 1.07 | 1.07 | 1.09 |
| MCR-AR-Gauss | 1 | 1.24 | 1.06 | 1.14 | 1.11 |
| MCR-ALS | 1 | 0.99 | 1.04 | 1.18 | 1.11 |
| Truncated SVD | 1 | 1.01 | 1.02 | 1.03 | 1.14 |
| MCR-ALS-random | 1 | 1.14 | 1.24 | 1.27 | 1.27 |
| MCR-NNLS-random | 1 | 1.17 | 1.19 | 1.29 | 1.38 |
| SVD | 1 | 1.37 | 1.42 | 1.44 | 1.46 |
| MCR-AR-Gauss-random | 1 | 1.20 | 1.25 | 1.36 | 1.61 |
| SOBI | 1 | 1.42 | 1.57 | 1.64 | 1.67 |
| MILCA | 1 | 1.25 | 1.52 | 1.77 | 1.97 |
| SIMPLISMA | 1 | 1.91 | 2.39 | 2.65 | 2.84 |
| FastICA | 1 | 1.72 | 2.39 | 3.27 | 3.40 |

Sub techniques are grouped together with the exception of MCR. MCR demonstrated varied results depending on the sub technique.

Our dataset can be separated into six groups of increasing noise level. We report the mean of the mean squared error for each general technique for predictions at the exact number of components as the noise level increases in Table 3, and present this information graphically in Figure 3. The best six techniques appear to show stable to mildly decreasing performance despite the increasing noise level.

Table 3. The mean of mean squared errors for each technique at increasing noise level.

| Technique[1] | Noise Factor | | | | | |
|---|---|---|---|---|---|---|
| | 0 | 0.0001 | 0.000178 | 0.000316 | 0.000562 | 0.001 |
| FastICA | 0.0 | 0.1 | 0.2 | 0.4 | 0.3 | 0.3 |
| SIMPLISMA | 0.2 | 0.3 | 0.3 | 0.5 | 0.4 | 0.4 |
| NNMF | 0.7 | 0.2 | 0.3 | 0.4 | 0.3 | 0.4 |
| TGA | 0.4 | 0.4 | 0.4 | 0.5 | 0.4 | 0.5 |
| VCA | 0.5 | 0.5 | 0.3 | 0.4 | 0.4 | 0.6 |
| JADE | 0.5 | 0.4 | 0.4 | 0.4 | 0.4 | 0.6 |
| MILCA | 0.2 | 0.6 | 0.3 | 0.6 | 0.5 | 0.8 |
| PARAFAC | 0.4 | 0.6 | 0.5 | 0.6 | 0.5 | 1.0 |
| MCR[2] | 0.9 | 0.6 | 0.5 | 0.6 | 0.6 | 0.7 |
| Truncated SVD | 0.5 | 0.7 | 0.7 | 0.8 | 0.6 | 1.4 |
| PCA | 0.5 | 0.7 | 0.7 | 0.8 | 0.6 | 1.5 |
| Incremental PCA | 0.5 | 0.7 | 0.7 | 0.8 | 0.7 | 1.5 |
| Sobi | 0.5 | 0.9 | 0.9 | 0.9 | 1.0 | 1.1 |
| Sparse PCA | 0.8 | 0.9 | 0.9 | 0.8 | 0.9 | 1.1 |
| SVD | 2.3 | 0.8 | 0.7 | 0.9 | 0.7 | 1.1 |

Noise factor is further explained in the methods section and represents a value multiplied to Gaussian noise added to the spectra. [1]Techniques are ordered by the mean of the mean squared error at all noise levels, which is equal to the value listed in table 1, No Normalization. [2]The MCR gaussian approach is excluded from this summary.

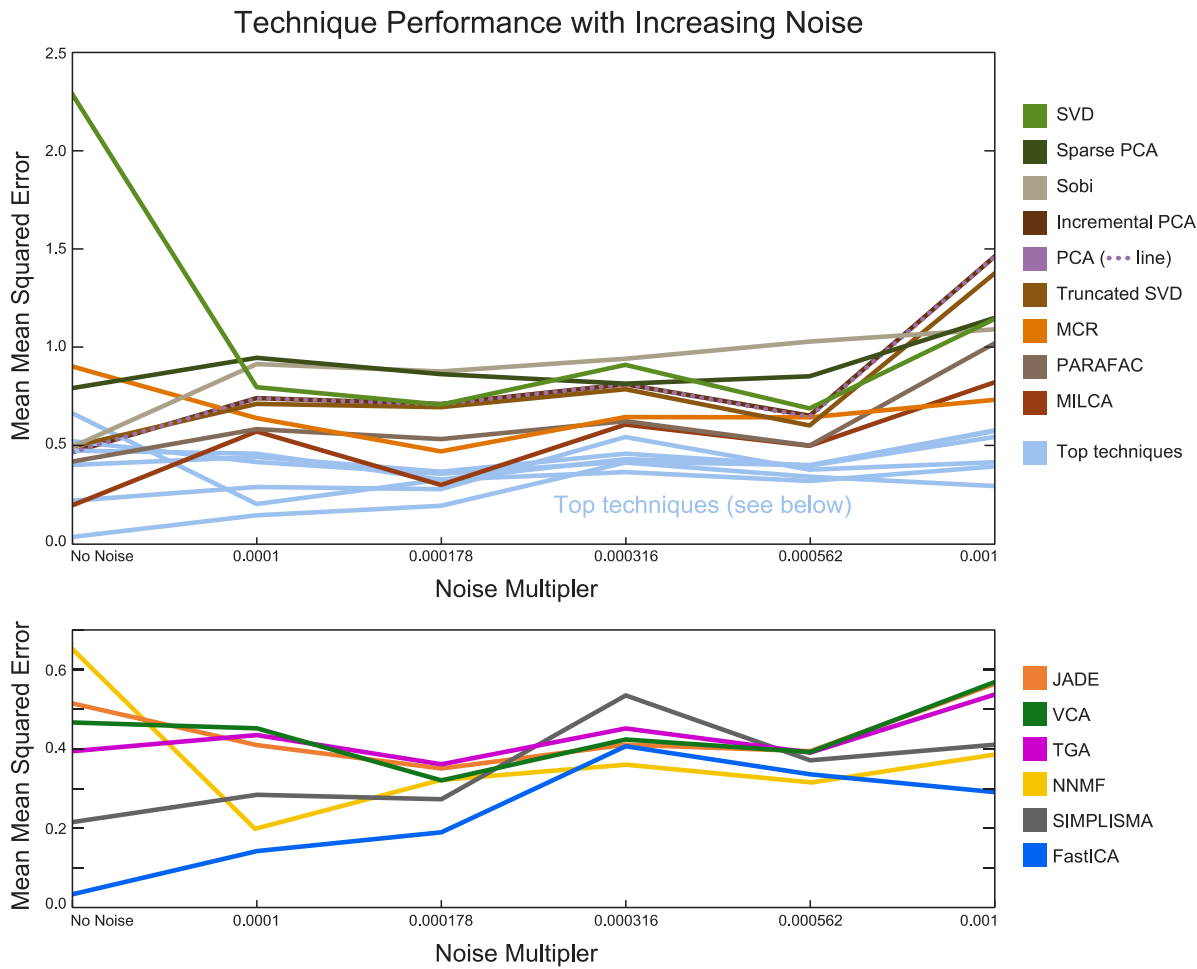

Figure 3. The performance of each technique with increasing noise. The top six best preforming techniques are detailed in the lower plot. Only "no normalization" values are plotted. Sub techniques (such as the various offset values for SIMPLISMA) were averaged together. The plotted values relate to those from Table 3.

3.2. Results from experimental datasets

3.2.1. Performance on real $^1$H MAS NMR $T_1$ datasets

We applied our benchmarked techniques to real solid state $^1$H MAS NMR $T_1$ spectral datasets result from mixtures of two chemical species. High-quality spectra of isolated molecules were collected for comparison. We applied the blind source separation techniques and then used our same appraisal method used on the synthetic datasets. Figure 4 and 5, depict the dataset and the most accurate prediction. It should be noted that possible differential spin-spin relaxation times ($T_2$) between the various chemical species would need to be addressed to provide accurate quantification of the relative concentrations of the components in the mixture. This can be done by obtaining NMR spectra for different Hahn echo times ($2\tau$), determining the concentration of each component in the mixture, and then back extrapolating to obtain the $\tau = 0$ concentration. While this was not performed in the present analysis, collecting additional spectra and using the predicted components could be quickly performed to produce accurate quantification the chemical species concentrations. Table 4 ranks the techniques as they performed on the two $^1$H MAS NMR $T_1$ datasets and visualizes the ranking in terms of relative mean squared error.

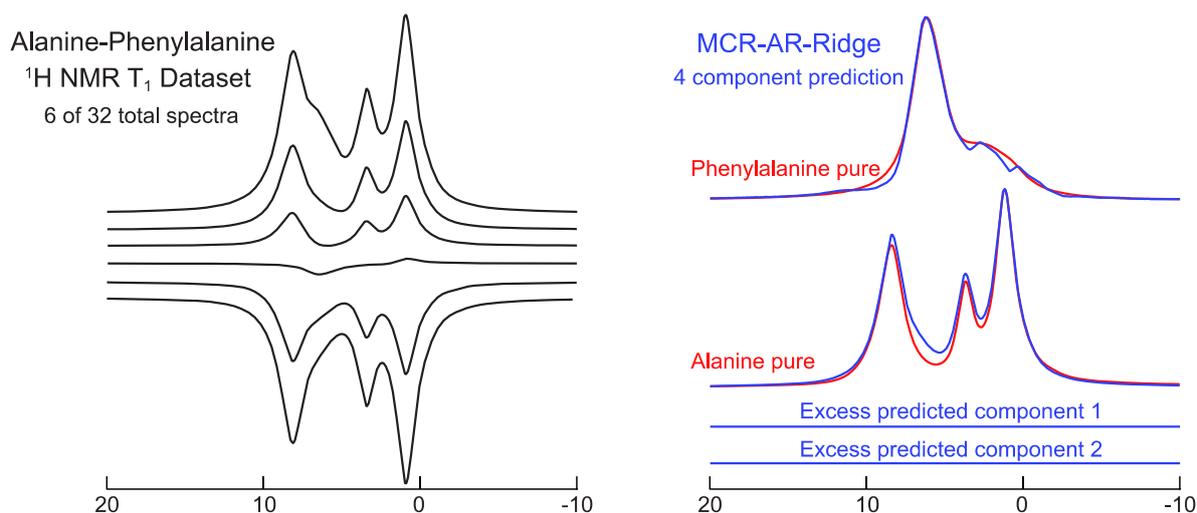

Figure 4. The $^1$H MAS NMR inversion recovery $T_1$ dataset containing alanine and approximately 20% phenylalanine. The most accurate prediction of the tested techniques was MCR-AR using ridge regression, and overpredicting 2 excess components. The pure component of phenylalanine and alanine are drawn on the right in red. The predicted components are superimposed over the pure components. If the prediction were perfect, no red would be visible. The excess predicted components contain no intensity in the pictured window, but appear to fit minor differences in baseline at the edges of the data.

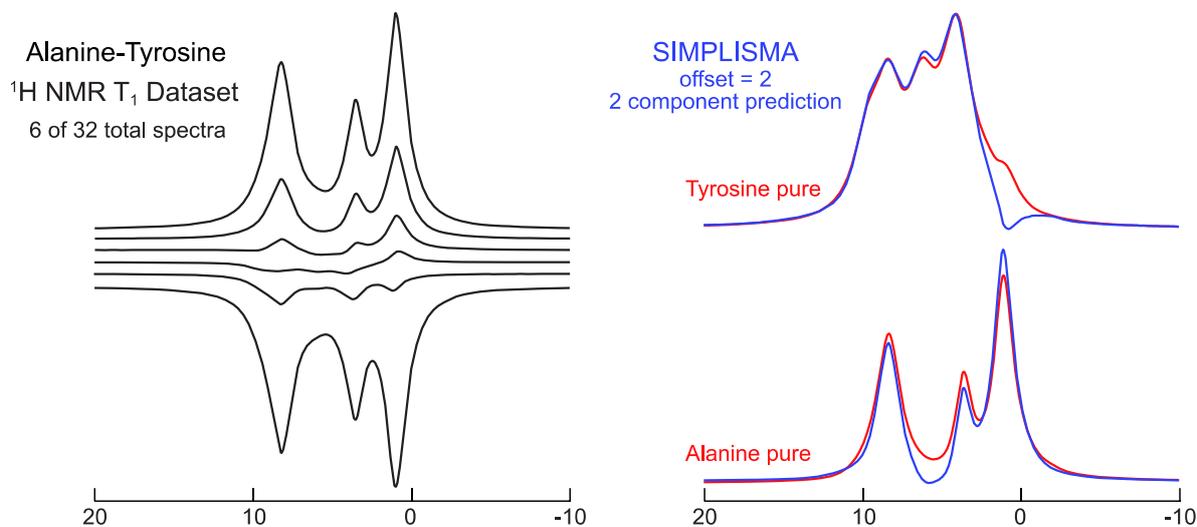

Figure 5. A figure of the $^1$H MAS NMR inversion recovery $T_1$ dataset containing alanine and a small fraction of tyrosine. The most accurate prediction of the tested techniques was SIMPLISMA, which picture on the left. The pure component of tyrosine and alanine are drawn on the right in red. The predicted components are superimposed over the pure components. If the prediction were perfect, no red would be visible.

Table 4. Mean of the mean squared error for each technique applied to both real $^1$H MAS NMR datasets.

| Technique | Relative Mean Squared Error |
|---:|:---:|
| SIMPLISMA-offset2 | 1.0 |
| NNMF-nndsvd | 4.3 |
| NNMF-random | 6.0 |
| MCR-NNLS | 6.9 |
| NNMF-nndsvdar | 7.2 |
| PARAFAC-random | 7.6 |
| NNMF-nndsvda | 8.3 |
| SIMPLISMA-offset0 | 8.5 |
| MCR-AR-Gauss | 9.1 |
| VCA | 9.6 |
| MCR-AR-Ridge-random | 9.9 |
| SIMPLISMA-offset15 | 11.7 |
| SIMPLISMA-offset8 | 11.8 |
| SIMPLISMA-offset12 | 13.1 |
| Incremental PCA | 14.0 |
| PCA | 14.0 |
| FastICA | 14.1 |
| Sparse PCA-cd | 14.3 |
| Sparse PCA-lars | 14.3 |
| Truncated SVD-arpack | 14.3 |
| Truncated SVD-randomized | 14.3 |
| PARAFAC-svd | 14.3 |
| MCR-AR-Gauss-random | 15.6 |
| MILCA-cd | 16.9 |
| MILCA-lars | 16.9 |
| TGA | 19.4 |
| MCR-AR-Ridge | 22.0 |
| MCR-ALS-random | 22.0 |
| MCR-ALS | 22.0 |
| SVD | 22.0 |
| MCR-NNLS-random | 22.5 |
| SOBI | 28.1 |
| JADE | 37.9 |

Mean squared errors are normalized relative to the best technique. Mean squared errors are the sum of the best prediction from each technique.

Several blind source techniques produced very accurate predictions, which are colored blue in the table and comprise approximately 1/3 of the tested techniques. There is a grouping of techniques that demonstrate midrange performance and a later grouping of poorly performing techniques.

Despite the relatively good performance of some techniques, there are still some minor differences across the pure and predicted components, as visible by the red lines in figures 4 and 5.

3.2.2. $^{27}$Al NMR nutation dataset

Determining the best technique and most accurate predicted components can be challenging when working in largely unknown systems. To demonstrate the realistic challenges and potential value and limits of using blind source separation on datasets without a known answer, we applied the best techniques to a solid-state aluminum $^{27}$Al NMR dataset of solid-state lasing rod material Thulium ($Tm^{3+}$) doped yttrium aluminum garnet, which presents overlapping paramagnetically shifted $AlO_6$ and $AlO_4$ peaks. $^{27}$Al is a quadrupolar nuclei (spin I = 5/2) with the magnitude of the quadrupolar interaction depending on the local bonding environments of the Al ($AlO_4$, $AlO_5$, $AlO_6$ etc). When this material is not paramagnetic, the spectra are easy to interpret, with two clear resonances relating to an $AlO_4$ coordination environment and an $AlO_6$ environment. The addition of $Tm^{3+}$ (paramagnetic with 2 unpaired electron spins) to the material shifts peak intensity from its diamagnetic isotopic chemical shift, resulting in peak broadening and the appearance of additional peaks. This nutation NMR dataset was originally used to identify the location of paramagnetically shifted peaks related to the $AlO_6$ site, but clear identification of shifted $AlO_4$ peaks was unobtainable. The dataset contains at least 2 components, one related to the $AlO_6$ site, and the other to the $AlO_4$ site, both with multiple peaks and sidebands overlapping in the region studied. Additional components could originate from differences between the central and satellite quadrupole transitions, or due to the distribution and local structure of doping lanthanide metal.

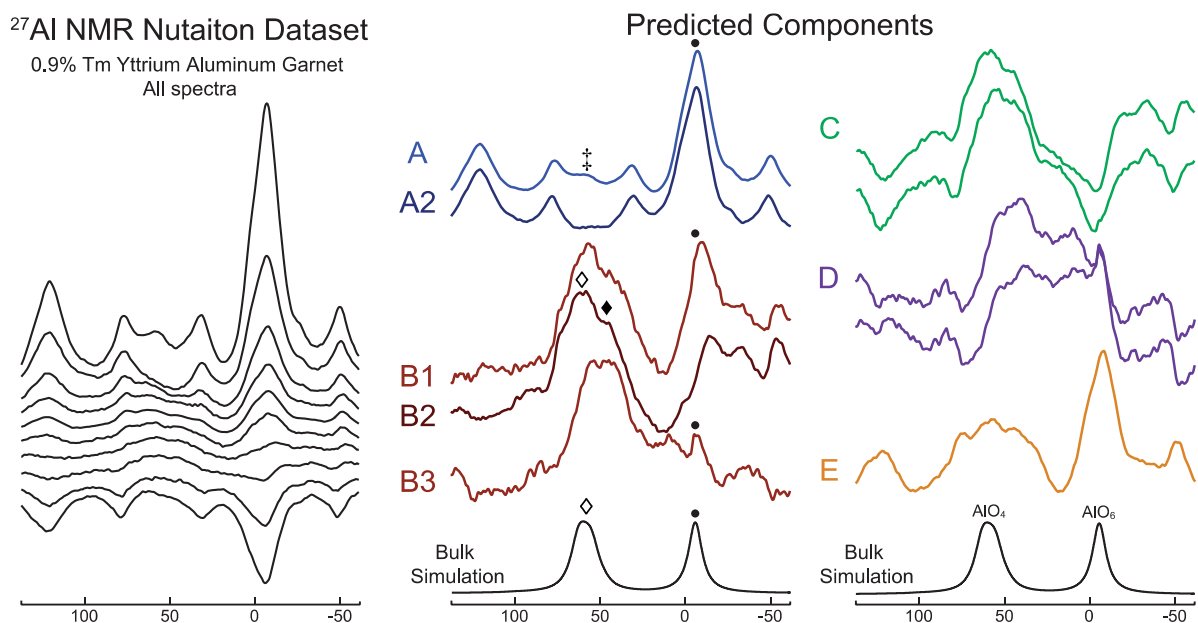

Figure 6. The $^{27}$Al NMR nutation spectra from [29], and predicted components. All 10 spectra are depicted on the left side. Examples of predicted components are depicted on the right side, as well as a simulation of the bulk sample (resulting from the majority of Al cations, which are structurally distant to the paramagnetic dopants but still show some paramagnetic broadening and shift). The letters refer to Table 5, and indicate which techniques produce components that qualitatively resemble these examples.

Table 5. A summary of common components predicted from the $^{27}$Al NMR nutation dataset

| Technique | 1 predicted component | 2 predicted components | 3 predicted components |
|---|---|---|---|
| FastICA | A | A, B2 | A, B2, D |
| JADE | A | A, A | A, A2, B1 |
| MCR-AR-Ridge-random | A-t | A-t, B2-t | A-t, B1-t, noise |
| MCR-NNLS-random | A-t | A-t, A-t | A-t, A-t, B1-t |
| NNMF-nndsvd | A | A, B1 | A, B1, B,1 |
| NNMF-nndsvda | A | artifacts | A2, B1-t, B1-t |
| NNMF-nndsvdar | A | A, B1 | A1, B1, B1 |
| NNMF-random | A | A, B1 | A2, B1, B2 |
| PARAFAC-random | A | A, A-B mixed | A, A2, A2 |
| SIMPLISMA-offset0 | A | A, B3 | A, B2, E |
| SIMPLISMA-offset12 | A | A, A2 | A, B1, C |
| SIMPLISMA-offset15 | A2 | A, A | A, A2, D |
| SIMPLISMA-offset2 | A | A, B3 | A2, A2, E |
| SIMPLISMA-offset8 | A | A, A | A2, A2, E |
| TGA | A | A, B2 | A, B2, D |
| VCA | A | B2, B2 | A2, B2, C |

Refer to Figure 6 for a depiction of the predicted components. "-t" indicates truncated baselines or peaks of components. "noise" indicates a component with no discernable peaks or features. "artifacts" indicates a component with large single-pixel spike-like artifacts.

The top two predicted components, A and B, are related to the AlO$_6$ site and the AlO$_4$ site respectively. The A peak and A2 peak are substantially similar with the exception of the small peak (marked with a ‡ in Figure 6). The agreement of these two features and the prevalence of an A-like features from each method gives strong evidence that the many peaks in A originate from the same component. The disagreement between the two components indicates that the inclusion of the ‡ peak as an AlO$_6$ feature is uncertain. If the origin of the ‡ peak was chemically important, a similar study to this one at a different spinning speed, higher magnetic field, different sample temperature, or by incorporating a wider spectrum window would likely confirm which pure component (A or A2) is most accurate. The identification of the B component highlights the usefulness of blind source separation. The input data (the experimental $^{27}$Al NMR Nutation Dataset in Figure 6) was intentionally collected to minimize the AlO$_4$ feature, and is difficult, near impossible, to visually identify in the dataset. Furthermore, the left-most peak appears to be a combination of two peaks with identical peak width (marked in figure 6 with a filled and hollow diamond, ◊,♦), one of which can be easily produced by broadening the AlO$_4$ at its expected location, and a neighboring peak which could be explained by a small paramagnetic shift. The similar location of the ‡ peak in component A, and the expected AlO$_4$ peak (◊) adds to the uncertainty. A common inaccuracy in blind source separation is the mixing of pure components, an example of which is visible in the B3 and marked with a ●. The C D and E components hint at the possibility of additional paramagnetic shifts that were unidentified in the original study. C, D and E also indicate the limits of blind source separation. With the given dataset size, quality of the spectra, and the inconsistency of the C, D and E predictions, the accuracy of these later components should be view with skepticism. The C, D and E predictions do suggest possible peaks that could be investigated and confirmed or ruled out using additional data, interpretation, and chemical constraints.

When attempting to interpret a large number of predicted components (96 total: 16 methods and 1, 2 and 3 components), we found it useful to first group the predictions into visually similar groups. Naming the components and creating the table and figure while interpreting the results was useful in interpreting and understanding the results. If we were studying the material and spectra above, the next steps would be to use traditional peak modeling and fitting constrained by reasonable quadrupolar parameters, incorporating both the blind source separation results and the original spectra to produce a model. Resonances visible in some of the predicted components, such as B, C and D, could likely be ruled out or assigned chemical meaning by collecting an additional dataset at a higher magnetic field, different radio frequency nutation field, or at a different spinning speed.

*3.3. Computational performance of our testing framework*

Using the functions modified from ssNake [27], generating and saving 32,000 unique pure components took less than 5 core hours @ 3.7 GHz, considerably faster than other NMR peak modeling techniques we have used in the past. Generating mixture datasets from pure components took ~2 core minutes @ 3.7 GHz for ~400 mixture spectra. Testing the blind source separation techniques on our X datasets, containing a total of Y spectra took ~6000 core hours @ 3.7 GHz (3 weeks X 7 days X 24 hr X 12 cores). Creating the optimized matched ensembles of pure to predicted components took ~1000 core hours @ 3.7 GHz, the majority of the time spent on the mixtures with 7, 8, 9 and 10 pure components.

4. Discussion

*4.1. General performance*

The mean of the mean squared error of all the blind source techniques explored are within 1 order of magnitude of each other and indicates that the most accurate technique is FastICA. However, the relatively similar mean squared errors stresses that including other performance metrics (e.g., minimum and variance of mean squared error) when selecting the best technique is important. SIMPLISMA has the capacity to produce remarkably exact pure components, as evident in Figure 2 and demonstrated in Figure 5. NNMF also stands out as a good technique in this regard. MILCA, a technique that otherwise does not seem particularly noteworthy, also produced accurate predicted components. As noise level increases, FastICA, SIMPLISMA, and NNMF again show the best performance. Under conditions of overprediction, where too many components were used during prediction, about half of the techniques (VCA, MCR-NNLS, PCA, Incremental PCA, Sparse PCA, TGA, PARAFAC, JADE, and NNMF) show stable performance. Caution should be used when employing the other techniques in systems where the number of pure components cannot be independently determined. On our small real data test set, SIMPLISMA, NNMF, MCR and PARAFAC performed well. Overall, our characterization of performance supports FastICA, SIMPLISMA, and NNMF as top-performing blind source prediction techniques.

Including the MCR-AR-Gauss method set an intriguing point of comparison. The method creates a single Gaussian peak per component, but within these constraints, it does a good job of minimizing the total error. When included peaks are not Gaussian (i.e. $2^{nd}$ order quadrupolar line shape), resulting predictions can be comically wrong (a Gaussian peak fit to a quadrupolar peak) but the resulting errors are often competitive to the other blind source separation techniques. This seems to imply that techniques with similar or higher mean squared errors are performing poorly.

Assessing technique performance on the inversion and nutation dataset supports a similar ranking scheme to the synthetic dataset. The nutation dataset revealed superior algorithm performance. This can be explained by the larger variation between the intensity of the different components, a product of the dataset's construction. This serves as a good reminder that the larger the variation between the individual components, the higher the likelihood of being able to identify individual components.

Common errors encountered in predicted components include mixed pure components and splitting of components. Mixed pure components consist of two or more pure components represented in a predicted component, often with opposite sign. An example of this effect is visible in Figure 4, where the large peak of alanine results in a small reduction of tyrosine's intensity. Split components occur when the combination of multiple predicted components is needed to produce a single pure component. This error seemed less common in the synthetic dataset than we anticipated, but was most frequently encountered when overpredicting components with the lower-ranked techniques of Table 2.

These techniques, although tested on data representative of NMR results, should be expected to transfer well to other spectroscopy datasets containing negative components, and we suspect that much of this is relevant to exclusively positive spectra. Our synthetic datasets containing quadrupolar peaks should be directly transferable to error analysis Gaussian peaks datasets, with the exception of the MCR-AR-Gauss methods, that we expect to demonstrate improved performance.

4.1.1. Performance relative to dataset normalization

The influence of normalization type, displayed in Table 1, can be useful in determining a choice of normalization. A large difference in the mean squared error between normalization types implies that the choice of normalization is important, whereas a small difference implies that this selection is unimportant. Several techniques, such as TGA, JADE, MILCA, Incremental PCA, Sparse PCA, and SVD, are not influenced by external normalization as they incorporate normalization within the algorithm. Our datasets are also constructed such that most spectra have related and somewhat similar intensity as is likely experienced when working with real datasets. We grouped techniques that were influenced by normalization into three categories, moderately influenced, minimally influenced, and negligibly influenced.

In general, our synthetic results seem to indicate that no normalization should be applied to the input data. SIMPLISMA is moderately influenced by normalization, and the best performance originates from the dataset without normalization. MCR is moderately to minimally influenced by normalization. However, the selection of normalization varies. When using ridge regression (MCR-AR-Ridge), a peak area normalization results in the lowest mean squared error and alternating least squares (MCR-ALS) prefers no normalization. MCR-AR-Gauss and MCR-NNLS methods do not have a consistent choice and are split between no normalization and normalization on by peak maximum. FastICA, PARAFAC-random, NNMF, VCA, and SOBI are minimally influenced by normalization type. PCA and NNMF-nndsvdar, NNMF-nndsvd, truncated SVD methods, and PARAFAC-svd are so minimally influenced by normalization, that the choice of normalization is unimportant. This finding is an expected outcome from the mathematical construction of PCA or the incorporation of SVD initialization. We were expecting drastic differences in terms of normalization from our personal experience in applying blind source separation techniques; however, on average, our study does not support this. These values may be useful in future studies as an initial hypothesis for determining the statistical significance of normalization results.

4.1.2. Overprediction of components

The correct number of components to predict can be challenging to determine in fully unknown systems. Table 2 gives some indication as to the risk and algorithm response when incorrectly choosing the number of components. It is desirable that when predicting excess (unreal) components, the previously predicted real components will not be degraded or changed. The majority of the techniques perform considerably well at maintaining accurate predictions when predicting excess components. It is important to remember that our approach for appraising the accuracy of a technique will discard predicted components which result in poor matches when there are excess predictions. Meaningful implementation of table 2's results in unknown systems requires that the human user can identify and discard excess components when using these techniques.

As the number of components predicted increases beyond the real number of components, VCA, MCR-NNLS, PCA, Incremental PCA, Sparse PCA, TGA do not degrade the accuracy of previously predicted components. Overprediction with these techniques does not have any additional drawbacks besides having excess components. Spectroscopists familiar with their systems can often identify excess components that are unrealistic and discard them from their analysis. The overall quality of predicted components is minimally degraded when overpredicting with PARAFAC, JADE, NNMF. MCR-AR-Ridge, MCR-AR-Gauss, MCR-ALS, and Truncated SVD show some disadvantage when overpredicting, and SOBI, MCR approaches using random initialization conditions, MILCA, and SVD result in poorer quality fits when predicting too many components. The top two techniques in terms of accuracy, FastICA and SIMPLISMA, perform poorly when overpredicting additional components.

In our experience, when working with real datasets, predicting additional components can often be useful, as "real" components are often unintentionally preset in the datasets. Examples of unintentional components include components that correct for inaccuracies in phasing or background signal, which reproducibly originate from the spectrometer. When using FastICA, SIMPLISMA, SOBI, MILCA, SVD, or MCR, practitioners should be wary about overpredicting components. The alanine-phenylalanine dataset presents an example of advantageous overprediction. Minor differences in baseline correction at the edges of the spectral window (and not pictured in Figure 4) were fit with excess predicted components, and the pure components were predicted most accurately. These two excess components are flat in the chemically relevant region of the spectra and would be easy for a human user to exclude from further analysis.

4.1.3. Experimental Results

The experimental results seem to agree with our synthetic benchmarks. Figures 4 and 5 indicate how accurate and potentially useful blind source separation can be, and show great performance for using the approach on NMR datasets with negative intensity. The close match on the central peaks as well as the spinning sidebands illustrate the effectiveness of blind source techniques at separating large and small related components.

The ranking produced on the two experimental datasets is similar to the ranking produced on the synthetic dataset. JADE and TGA are an exception to the similar performance, as they perform poorly and drop significantly in ranking. The synthetic dataset ranking is based on the averaging values, and as indicated in Figure 2, bad results from even the best techniques are expected some of the time. A fundamental challenge when using blind source separation techniques is determining if a good approach is producing unreasonable results.

The results in Table 4 may have implications for blind source separation practitioners. On only two very similar datasets, the techniques' performance varied widely. Great accuracy was produced in some cases with dismal performance in others. The variety of results seems to indicate that for a given dataset, it is likely that some technique will outperform the others. This implies that the traditional chemometric approach of using the results from only one or two algorithms misses the opportunity to identify much more accurate predictions. Broader consideration of the results from an ensemble of techniques for each dataset has the potential of connecting users with useful insight.

With the $^{27}$Al NMR nutation dataset, we used multiple blind source separation techniques to begin to build an interpretation of the data. This approach produces many predicted components for the user to interpret but does provide information that is unrealistic to decern directly from the dataset. Incorporating other chemical insight, such as physical constraints from the chemical system, as well as other modeling techniques and NMR peak fitting, could likely help identify which of the predicted components are most reasonable.

*4.2. Recommended techniques for matrix initialization*

Initializing a starting matrix, especially when the number of components is not known, VCA appears to be an ideal choice. If the number of components is known exactly, FastICA and SIMPLISMA are better alternatives, and when the real number of components is not known, TGA or JADE appear to be acceptable alternatives to VCA. Typically, if a blind source separation practitioner is initializing a matrix, the number of components is likely not known. VCA, TGA, and JADE have comparable accuracy on the synthetic dataset, run times within the same order of magnitude on datasets our size, and maintain the accuracy of predicted components when overpredicting.

In other fields, the standard choice for matrix initialization is SVD (as we use in this study for initializing several techniques), but our work seems to indicate that this is not an advantageous choice for NMR-like data. There is some precedent for alternative initialization techniques for NMR spectra; Cherni et al. use the work of Toumi et al. to motivate the choice of JADE for initializing the starting matrix [18,20]. Although we also recommend JADE, an interesting drawback of JADE is its reliance on a SVD initialization, which for large datasets, can result in considerably slow performance. A directed study on matrix initialization for NMR datasets would be useful to confirm the VCA recommendation our data seems to support.

*4.3. Recommendations for collecting datasets*

While inspecting the synthetic and real datasets, we noticed that spectra near the inflection from negative to positive signal intensity often contained the largest degree of variance between components. Larger variance between components is typically advantageous for blind source separation techniques. This implies that the data near the inflection point is valuable for the separation, and when collecting data, additional or higher quality data close to the inversion point may be useful for increasing the accuracy of the predicted components.

*4.4. Drawbacks of the present study*

In our effort to understand the influence of noise on the techniques, we included synthetic datasets without any noise. Although most techniques applied to the no-noise spectra present a trend that agrees with increasing noise, MCR and SVD appear to have a particular challenge when working with no-noise datasets and the no-noise datasets somewhat unfairly skew their overall performance. Since no-noise datasets are unrealistic in the real-world settings, we recommend no-noise datasets be avoided in future work benchmarking techniques for application to spectroscopic data.

Additionally, we used only Gaussian noise, which does not reflect the full range of complications experienced in real-life experimental applications. Real applications do frequently include substantial Gaussian noise, but also many other issues such as baseline correction artifacts, phasing, acoustic probe ringing, digital filtering anomalies, and background signals. Future studies with an expanded amount of noise and additional types of noise would be useful.

Our experimental NMR test datasets contain only 2 (or likely 2) pure components. Although this enables a simple demonstration of blind source separation to negative intensity NMR spectra, it is somewhat removed from applications where blind source separation is needed: datasets with a large number of components that are challenging to address using more traditional spectroscopic and analysis techniques.

*4.5. Algorithm run times*

Although we reported a rough run time speed for each blind source separation technique, runtime is a nuanced characteristic. First and foremost, due to the relatively small volume of NMR data (often dictated by spectrometer time and the expense of samples), only in exceptional cases should the run time of the algorithms be an important practical consideration. Slower techniques are still sufficiently fast when dealing with a single experimental dataset. However, slow run time techniques do limit the ability of a scientist to check and optimize hyperparameters (such as the number of components or choice of spectra normalization), and limit rapid, hands-on, iterative testing which can be useful in helping the user understand the dataset and algorithm. The run time of matrix decomposition algorithms may not be transferable to other matrices, as discussed in Halko et al., as the properties of a matrix can vary greatly, and some matrix decomposition techniques are strongly influenced by the starting matrix [16].

Slow algorithms are limiting for benchmark studies. Several blind source separation techniques which we intended to report on became prohibitively expensive in terms of CPU time to benchmark on the large number of test spectra we used. This includes an alternative robust PCA approach, RADICAL (Robust, Accurate, Direct ICA aLgorithm), and Kernel PCA. Some of the algorithms we used would have been unrealistic for our project if the spectra contained additional points. In a few exploratory test cases, we tested larger synthetic spectra (c.a. 160,000 points instead of 10,000 or 1024), and did not notice a significant decline in algorithm accuracy, although the runtimes were substantially longer. Further work addressing the performance of windowed (cropped) spectra vs. large sweep widths would likely be useful for blind source separation practitioners. Other techniques that we were interested in benchmarking became prohibitively expensive in terms of the coding time needed to implement them. In a few cases, reproducibility of recently published algorithms for which source code was not provided was complicated by sparse methodical descriptions, unresponsiveness when contacted, or lost research materials.

*4.6. Our benchmarking framework*

The method we used for matching predicted components to pure components has several drawbacks. The combinatoric matching approach works well for a small number of components (1-6 or so), but as the number of components increases, the combinations that need to be considered grows exponentially. In its unmodified form, benchmarking datasets with more than 15 pure components become computationally unfeasible, and benchmarking datasets with more than 6 pure components can take considerable time (~2 or more hours). Determining goodness of fit using the lack-of-fit sum of squares seems to provide meaningful insight. However, it could discredit a possible match in a few realistic settings. For instance, a nearly perfect predicted component with a few extreme-valued points resulted in a large squared error to a pure component despite an otherwise close match. While not present in our dataset, datasets with extreme artifacts or large amounts of noise could produce these. More importantly, a human can easily identify these few points as erroneous, something our method could not do.

While our matched ensemble approach is unable to perform well under the above circumstances, it maintains some advantage over human appraisal. Our method for benchmarking techniques is entirely blind to predicted components containing contributions from two or more pure components. If predicted components contain aspects of two pure components, one of these pure components is lost entirely, and its contributions are counted as error. In reality, an observant scientist could infer extra meaning from such a predicted component, where negative (subtractive) features associated with a positive feature implies an inverse correlation, and two positive features imply correlation. In our experience, this additional insight can be valuable in real settings, and this extra information can be incorporated into a final model. When considering settings which have predicted too few, or too many components, our matched ensemble approach is unable to meaningfully incorporate this information as it discards extra unused predicted components, and does not account for errors resulting from too few components. In contrast to these disadvantages, our matched ensemble approach is very accurate at determining the best ensemble match, and is not confused or deluded by predicted components that appear to be significant but are the products of noise or overlaps (which can create realistic looking peaks that have no meaning).

5. Conclusions

FastICA, SIMPLISMA, and NNMF all have several desirable characteristics that support a recommendation as top choice techniques for blind source separation on NMR spectral datasets containing negative intensity. We arrived at this recommendation after testing 15 different blind source separation algorithms, some of which contained an opportunity for additional settings, resulting in 33 total techniques tested. We benchmarked these techniques on two large datasets of synthetically generated NMR spectra representing $T_1$ inversion or nutation experiments. The mean accuracy of the most accurate technique and of the least accurate technique are within one order of magnitude, implying similar performance on average. The techniques also show wide variation in accuracy. SIMPLISMA and NNMF have the potential of predicting the most accurate pure components. Poor performance of FastICA and SIMPLISMA is expected if additional components beyond the correct number of pure components are employed during prediction.

Because of the variation of techniques on synthetic NMR datasets and the results from our application of the techniques to experimental NMR data, we recommend incorporating predictions from multiple blind source separation techniques. Using multiple techniques has the potential of connecting the user with a more accurate prediction than an approach that considers only a single technique. Due to the similar appearance and related qualities of NMR data to other spectroscopic techniques, we anticipate the conclusions in this paper to extend to analysis of data originating from other spectroscopic methods.

Key recommendations for blind source separation practitioners working with NMR $T_1$ inversion or nutation data:
- FastICA, SIMPLISMA, and NNMF are top choice techniques.

- A comparison of predicted components from multiple techniques applied to the same dataset may be useful in identifying the best predicted components.
- Normalizing the input dataset is likely not necessary.
- SVD preforms poorly. To initialize starting matrices, a more accurate alternative to SVD may result in lower errors. We recommend FastICA or SIMPLISMA when there is a clear expectation of pure components within a dataset, or VCA for a less accurate but more reliable approach.
- Collecting data clustered around the positive-negative spectral intensity inversion point may help increase accuracy.

Author Contributions:

conceptualization, R.J.M.;
methodology, R.J.M., and N.R.;
software, R.J.M., N.R., and M.W.;
validation, T.M.A.;
formal analysis, R.J.M. and M.W.;
resources, R.J.M.;
data curation, R.J.M.;
writing—original draft preparation, R.J.M.;
writing—review and editing, R.J.M. and T.M.A.;
visualization, R.J.M., and M.W.;
project administration, R.J.M.;
[CRediT taxonomy](#) for the term explanation.

Funding: R.J.M. is thankful for support from the University of California President's Postdoctoral Fellowship Program. T.M.A. is supported by Sandia National Labs.

Acknowledgments: The authors thank Ryan Padilla (Vincit California) for his python assistance in the project's early stages, and Patrick J. McCarty (General Atomics Aeronautical Systems, Inc.) for data backup and parallel processing advice. The $^1$H MAS NMR portion of the work (T.M.A.) was performed at Sandia National Laboratories which is a multi-mission laboratory managed and operated by National Technology and Engineering Solutions of Sandia, LLC., a wholly owned subsidiary of Honeywell International, Inc., for the U.S. Department of Energy's National Nuclear Security Administration under contract DE-NA0003525.

Conflicts of Interest: The authors declare no financial conflict of interest, and declare no intellectual conflict of interest motivating algorithm comparison results towards a particular technique or approach.

Key Terms and Abbreviations

Pure component (The XY data from a single source, it is the signal of a single compound from the samples of interest)
Predicted component (An algorithm's estimation of the pure component)
Mixture spectra (Sets of X, Y data which contain a contribution from one or more pure components + noise)
Dataset (A collection of mixture spectra which are thought to share similar features)
Algorithm (any computational method which receives input data and returns an outcome)
Technique (the application of an algorithm to data, where the user must select options (such as: what data, algorithm-specific settings, and if the results are valid))

Blind Source Separation Acronyms:
  BSS         Blind Source Separation
  ICA         Independent Component Analysis
  JADE        Joint Approximate Diagonalization of Eigenmatrices

| | |
|---|---|
| MCR | Multivariate Curve Resolution |
| MILCA | Mutual Information Least dependent Component Analysis |
| NN | Naanaa and Nuzillard method |
| NNMF | Non-Negative Matrix Factorization |
| PARAFAC | Parallel Factor Analysis |
| PCA | Principal Component Analysis |
| SIMPLISMA | SIMPLe-to-use-Interactive Self-modeling Mixture Analysis |
| SOBI | Second Order Blind Identification |
| SVD | Singular Value Decomposition |
| TGA | Trimmed Grassmann Average, a PCA variation |
| VCA | Vertex Component Analysis |

Sub-technique Acronyms:

| | |
|---|---|
| ALS | Alternating Least Squares |
| AR | Alternating Regression |
| ARPACK | ARnoldi PACKage, a fortran software package |
| cd | Coordinate Decent method |
| lars | Least Angle Regression |
| nndsvd | Non-Negative Double Singular Value Decomposition |
| nndsvda | nndsvd with zero values replaced with the average point value of the input dataset |
| nndsvdar | nndsvd with zero values replaced with small random values |
| NNLS | Non-Negative Least Squares |

NMR Acronyms:

| | |
|---|---|
| NMR | Nuclear Magnetic Resonance |
| CP/MAS | Cross-Polarization Magic Angle Spinning, an NMR technique |
| TOCSY-$t_1$ | TOtal Correlation SpectroscopY |

References

1. Kofidis, E. Blind Source Separation: Fundamentals and Recent Advances. *arXiv* 2016, 1–46. arXiv: 1610.09555.

2. Toumi, I.; Caldarelli, S.; Torrésani, B. A review of blind source separation in NMR spectroscopy. *Prog. Nucl. Magn. Reson. Spectrosc.* 2014, *81*, 37–64, doi:10.1016/j.pnmrs.2014.06.002.

3. Alam, T.M.; Alam, M.K. Chemometric Analysis of NMR Spectroscopy Data: A Review. *Annu. Reports NMR Spectrosc.* 2004, *54*, 41–80, doi:10.1016/S0066-4103(04)54002-4.

4. Ebrahimi, P.; Viereck, N.; Bro, R.; Engelsen, S.B. Chemometric Analysis of NMR Spectra. In *Modern Magnetic Resonance*; Webb, G.A., Ed.; Springer International Publishing: Cham, 2016; pp. 1–20 ISBN 978-3-319-28275-6.

5. Windig, W.; Antalek, B. Direct exponential curve resolution algorithm (DECRA): A novel application of the generalized rank annihilation method for a single spectral mixture data set with exponentially decaying contribution profiles. *Chemom. Intell. Lab. Syst.* 1997, *37*, 241–254, doi:10.1016/S0169-7439(97)00028-2.

6. Windig, W.; Antalek, B. Resolving nuclear magnetic resonance data of complex mixtures by three-way methods: Examples of chemical solutions and the human brain. *Chemom. Intell. Lab. Syst.* 1999, *46*, 207–219, doi:10.1016/S0169-7439(98)00172-5.

7. Windig, W.; Antalek, B.; Sorriero, L.J.; Bijlsma, S.; Louwerse, D.J.; Smilde, A.K. Applications and new developments of the direct exponential curve resolution algorithm (DECRA). Examples of spectra and magnetic resonance images. *J. Chemom.* 1999, *13*, 95–110, doi:10.1002/(SICI)1099-128X(199903/04)13:2<95::AID-CEM530>3.0.CO;2-L.

8. Alam, T.M.; Alam, M.K. Effect of non-exponential and multi-exponential decay behavior on the performance of the direct exponential curve resolution algorithm (DECRA) in NMR investigations. *J. Chemom.* 2003, *17*, 583–593, doi:10.1002/cem.826.

9. Dal Poggetto, G.; Castanar, L.; Adams, R.W.; Morris, G.A.; Nilsson, M. Dissect and Divide: Putting NMR Spectra of Mixtures under the Knife. *J. Am. Chem. Soc.* 2019, *141*, 5766–5771, doi:10.1021/jacs.8b13290.

10. Monakhova, Y.B.; Ruge, W.; Kuballa, T.; Ilse, M.; Winkelmann, O.; Diehl, B.; Thomas, F.; Lachenmeier, D.W. Rapid approach to identify the presence of Arabica and Robusta species in coffee using $^1$H NMR spectroscopy. *Food Chem.* 2015, *182*, 178–184, doi:10.1016/j.foodchem.2015.02.132.

11. Yilmaz, A.; Nyberg, N.T.; Jaroszewski, J.W. Metabolic profiling based on two-dimensional J-resolved $^1$H NMR data and parallel factor analysis. *Anal. Chem.* 2011, *83*, 8278–8285, doi:10.1021/ac202089g.

12. Mason, H.E.; Uribe, E.C.; Shusterman, J.A. Rapid acquisition of data dense solid-state CPMG NMR spectral sets using multi-dimensional statistical analysis. *Phys. Chem. Chem. Phys.* 2018, *20*, 18082–18088, doi:10.1039/c8cp02382d.

13. Mason, H.E.; Begg, J.D.; Maxwell, R.S.; Kersting, A.B.; Zavarin, M. A novel solid-state NMR method for the investigation of trivalent lanthanide sorption on amorphous silica at low surface loadings. *Environ. Sci. Process. Impacts* 2016, *18*, 802–809, doi:10.1039/c6em00082g.

14. McCarty, R.J.; Stebbins, J.F. Constraints on aluminum and scandium substitution mechanisms in forsterite, periclase, and larnite: High-resolution NMR. *Am. Mineral.* 2017, *102*, 1244–1253, doi:10.2138/am-2017-5976.

15. Kusaka, Y.; Hasegawa, T.; Kaji, H. Noise Reduction in Solid-State NMR Spectra Using Principal Component Analysis. *J. Phys. Chem. A* 2019, doi:10.1021/acs.jpca.9b04437.

16. Halko, N.; Martinsson, P.G.; Tropp, J.A. Finding structure with randomness: Probabilistic algorithms for constructing approximate matrix decompositions. *SIAM Rev.* 2011, *53*, 217–288, doi:10.1137/090771806.

17. Monakhova, Y.B.; Tsikin, A.M.; Kuballa, T.; Lachenmeier, D.W.; Mushtakova, S.P. Independent component analysis (ICA) algorithms for improved spectral deconvolution of overlapped signals in $^1$H NMR analysis: application to foods and related products. *Magn. Reson. Chem.* 2014, *52*, 231–240, doi:10.1002/mrc.4059.

18. Toumi, I.; Torrésani, B.; Caldarelli, S. Effective processing of pulse field gradient NMR of mixtures by blind source separation. *Anal. Chem.* 2013, *85*, 11344–11351, doi:10.1021/ac402085x.

19. Naanaa, W.; Nuzillard, J.M. Blind source separation of positive and partially correlated data. *Signal Processing* 2005, *85*, 1711–1722, doi:10.1016/j.sigpro.2005.03.006.

20. Cherni, A.; Piersanti, E.; Anthoine, S.; Chaux, C.; Shintu, L.; Yemloul, M.; Torrésani, B. Challenges in the decomposition of 2D NMR spectra of mixtures of small molecules. *Faraday Discuss.* 2019, *218*, 459–480, doi:10.1039/c9fd00014c.

21. Python Software Foundation Python 2019.

22. Intel Corporation Intel® Distribution for Python 2019.

23. Van Der Walt, S.; Colbert, S.C.; Varoquaux, G. The NumPy array: A structure for efficient numerical computation. *Comput. Sci. Eng.* 2011, *13*, 22–30, doi:10.1109/MCSE.2011.37.

24. Lam, S.K.; Pitrou, A.; Seibert, S. Numba: A LLVM-based python JIT compiler. *Proc. Second Work. LLVM Compil. Infrastruct. HPC - LLVM '15* 2015, 1–6, doi:10.1145/2833157.2833162.

25. Hunter, J.D. Matplotlib: A 2D graphics environment. *Comput. Sci. Eng.* 2007, *9*, 99–104, doi:10.1109/MCSE.2007.55.

26. McKinney, W. Data Structures for Statistical Computing in Python. *Proc. 9th Python Sci. Conf.* 2010, 51–56.

27. van Meerten, S.G.J.; Franssen, W.M.J.; Kentgens, A.P.M. ssNake: A cross-platform open-source NMR data processing and fitting application. *J. Magn. Reson.* 2019, *301*, 56–66, doi:10.1016/j.jmr.2019.02.006.

28. Vitter, J.S. Random Sampling with a Reservoir. *ACM Trans. Math. Softw.* 1985, *11*, 37–57, doi:10.1145/3147.3165.

29. McCarty, R.J.; Stebbins, J.F. Investigating lanthanide dopant distributions in Yttrium Aluminum Garnet (YAG) using solid state paramagnetic NMR. *Solid State Nucl. Magn. Reson.* 2016, *79*, 11–22, doi:10.1016/j.ssnmr.2016.10.001.

30. Golub, G.H.; Reinsch, C. Singular value decomposition and least squares solutions. *Numer. Math.* 1970, *14*, 403–420, doi:10.1007/BF02163027.

31. Golub, G.; Kahan, W. Calculating the Singular Values and Pseudo-Inverse of a Matrix. *J. Soc. Ind. Appl. Math. Ser. B Numer. Anal.* 1965, *2*, 205–224, doi:10.1137/0702016.

32. Pedregosa, F.; Varoquaux, G.; Gramfort, A.; Michel, V.; Thirion, B.; Grisel, O.; Blondel, M.; Prettenhofer, P.; Weiss, R.; Dubourg, V.; et al. Scikit-learn: Machine Learning in Python. *J. Mach. Learn. Res.* 2011, *12*, 2825–2830.

33. Virtanen, P.; Gommers, R.; Oliphant, T.E.; Haberland, M.; Reddy, T.; Cournapeau, D.; Burovski, E.; Peterson, P.; Weckesser, W.; Bright, J.; et al. SciPy 1.0--Fundamental Algorithms for Scientific Computing in Python. *arXiv* 2019, 1–22. arXiv:1907.10121.

34. Pearson, K. LIII. On lines and planes of closest fit to systems of points in space. *London, Edinburgh, Dublin Philos. Mag. J. Sci.* 1901, *2*, 559–572, doi:10.1080/14786440109462720.

35. Hotelling, H. Analysis of a complex of statistical variables into principal components. *J. Educ. Psychol.* 1933, *24*, 417–441, doi:10.1037/h0071325.

36. Jenatton, R.; Obozinski, G.; Bach, F. Structured sparse principal component analysis. *J. Mach. Learn. Res.* 2010, *9*, 366–373.

37. Ross, D.A.; Lim, J.; Lin, R.S.; Yang, M.H. Incremental learning for robust visual tracking. *Int. J. Comput. Vis.* 2008, *77*, 125–141, doi:10.1007/s11263-007-0075-7.

38. Levey, A.; Lindenbaum, M. Sequential Karhunen-Loeve basis extraction and its application to images. *IEEE Trans. Image Process.* 2000, *9*, 1371–1374, doi:10.1109/83.855432.

39. Hauberg, S.; Feragen, A.; Black, M.J. Grassmann averages for scalable robust PCA. *Proc. IEEE Comput. Soc. Conf. Comput. Vis. Pattern Recognit.* 2014, 3810–3817, doi:10.1109/CVPR.2014.481.

40. Carroll, J.D.; Chang, J.J. Analysis of individual differences in multidimensional scaling via an n-way generalization of "Eckart-Young" decomposition. *Psychometrika* 1970, *35*, 283–319, doi:10.1007/BF02310791.

41. Harshman, R.A. Foundations of the PARAFAC procedure: Models and conditions for an "explanatory" multimodal factor analysis. *UCLA Working Papers in Phonetics* 1970, *16*, 1–84.

42. Kossaifi, J.; Panagakis, Y.; Anandkumar, A.; Pantic, M. TensorLy: Tensor learning in python. *J. Mach. Learn. Res.* 2019, *20*, 1–5. https://github.com/tensorly/tensorly. arXiv:1610.09555v2.

43. Bro, R. PARAFAC. Tutorial and applications. *Chemom. Intell. Lab. Syst.* 1997, *38*, 149–171, doi:10.1016/S0169-7439(97)00032-4.

44. Stögbauer, H.; Kraskov, A.; Astakhov, S.A.; Grassberger, P. Least-dependent-component analysis based on mutual information. *Phys. Rev. E - Stat. Physics, Plasmas, Fluids, Relat. Interdiscip. Top.* 2004, *70*, 17, doi:10.1103/PhysRevE.70.066123.

45. Kraskov, A.; Stögbauer, H.; Grassberger, P. Estimating mutual information. *Phys. Rev. E - Stat. Physics, Plasmas, Fluids, Relat. Interdiscip. Top.* 2004, *69*, 16, doi:10.1103/PhysRevE.69.066138.

46. Cardoso, J.F.; Souloumiac, A. Blind beamforming for non-Gaussian signals. *IEE Proceedings, Part F Radar Signal Process.* 1993, *140*, 362–370, doi:10.1049/ip-f-2.1993.0054.

47. Beckers, G.J.L. jadeR.py Available online: http://www.gbeckers.nl/pages/numpy_scripts/jadeR.py.

48. Cardoso, J.F. High-order contrasts for independent component analysis. *Neural Comput.* 1999, *11*, 157–192, doi:10.1162/089976699300016863.

49. Nascimento, J.M.P.; Dias, J.M.B. Vertex component analysis: a fast algorithm to unmix hyperspectral

data. *IEEE Trans. Geosci. Remote Sens.* 2005, *43*, 898–910, doi:10.1109/TGRS.2005.844293.

50. Lagrange, A. Vertex Component Analysis 2018. https://github.com/Laadr/VCA.

51. Lee, D.D.; Seung, H.S. Learning the parts of objects by non-negative matrix factorization. *Nature* 1999, *401*, 788–791, doi:10.1038/44565.

52. Cichocki, A.; Phan, A.H. Fast local algorithms for large scale nonnegative matrix and tensor factorizations. *IEICE Trans. Fundam. Electron. Commun. Comput. Sci.* 2009, *E92-A*, 708–721, doi:10.1587/transfun.E92.A.708.

53. Boutsidis, C.; Gallopoulos, E. SVD based initialization: A head start for nonnegative matrix factorization. *Pattern Recognit.* 2008, *41*, 1350–1362, doi:10.1016/j.patcog.2007.09.010.

54. Belouchrani, A.; Abed-Meraim, K.; Cardoso, J.F.; Moulines, E. A blind source separation technique using second-order statistics. *IEEE Trans. Signal Process.* 1997, *45*, 434–444, doi:10.1109/78.554307.

55. Rigie, D. Second Order Blind Identification 2018. https://github.com/davidrigie/sobi.

56. Lawton, W.H.; Sylvestre, E.A. Self Modeling Curve Resolution. *Technometrics* 1971, *13*, 617, doi:10.2307/1267173.

57. Camp, C.H. pyMCR: A Python Library for MultivariateCurve Resolution Analysis with Alternating Regression (MCR-AR). *J. Res. Natl. Inst. Stand. Technol.* 2019, *124*, 1–10, doi:10.6028/jres.124.018.

58. Camp, C.H. pyMCR. *U.S. Deptment of Commerce, National Institute of Standards and Technology.* 2019, doi:10.18434/M32064.

59. Lawson, C.L.; Hanson, R.J. *Solving least squares problems*; SIAM, 1995; Vol. 15;.

60. Windig, W.; Guilment, J. Interactive self-modeling mixture analysis. *Anal. Chem.* 1991, *63*, 1425–1432, doi:10.1021/ac00014a016.

61. Windig, W. Spectral data files for self-modeling curve resolution with examples using the Simplisma approach. *Chemom. Intell. Lab. Syst.* 1997, *36*, 3–16, doi:10.1016/S0169-7439(96)00061-5.

62. Windig, W.; Antalek, B.; Lippert, J.L.; Batonneau, Y.; Brémard, C. Combined Use of Conventional and Second-Derivative Data in the SIMPLISMA Self-Modeling Mixture Analysis Approach. *Anal. Chem.* 2002, *74*, 1371–1379, doi:10.1021/ac0110911.

63. Nelder, J.A.; Mead, R. A Simplex Method for Function Minimization. *Comput. J.* 1965, *7*, 308–313, doi:10.1093/comjnl/7.4.308.